\documentclass{acm_proc_article-sp}

\usepackage{cite}
\usepackage{xspace}
\usepackage{url}
\usepackage{balance}

\newcommand{\impact}[4]{\textsf{\small [#1$\,\vert\,$\uppercase{{\scriptsize#2}}] 
$\stackrel{#4}{\longrightarrow}$ [#3]}\xspace}

\newcommand{\impactp}[3]{\impact{#1}{#2}{#3}{+}}
\newcommand{\impactn}[3]{\impact{#1}{#2}{#3}{-}}

\newcommand{\fact}[2]{\textsf{\small [#1$\,\vert\,$\uppercase{{\scriptsize#2}}]}}
\newcommand{\entity}[1]{\textsf{\small #1}}
\newcommand{\attribute}[1]{\textsf{\small\uppercase{{\scriptsize#1}}}}

\begin{document}
%

\title{A Bayesian Network Approach to Assess and Predict Software Quality Using Activity-Based
 Quality Models}

\numberofauthors{1}

\author{
  \alignauthor Stefan Wagner\\
    \affaddr{Fakult\"at f\"ur Informatik}\\
    \affaddr{Technische Universit\"at M\"unchen}\\
    \affaddr{Garching b.\ M\"unchen, Germany}\\
    \email{wagnerst@in.tum.de}
}


\maketitle

\begin{abstract}
Assessing and predicting the complex concept of software quality is still
challenging in practice as well as research. Activity-based quality models
break down this complex concept into more concrete definitions, more precisely
facts about the system, process and environment and their impact on
activities performed on and with the system. However, these models
lack an operationalisation that allows to use them in assessment
and prediction of quality. Bayesian Networks (BN) have been
shown to be a viable means for assessment and prediction incorporating
variables with uncertainty. This paper describes how activity-based quality
models can be used to derive BN models for quality assessment and
prediction. The proposed approach is demonstrated in a proof of concept
using publicly available data.
\end{abstract}

\category{D.2.8}{Software Engineering}{Metrics}
\category{D.2.9}{Software Engineering}{Management}[Software\ Quality
Assurance (SQA)]

\terms{Measurement, Management}

\keywords{Activity-Based Quality Models, Bayesian Networks, Quality Assessment,
Quality Prediction}

\section{Introduction\label{sec:intro}}

Despite the importance of software quality, the management of software quality
is still an immature discipline in software engineering research and practice.
Research work has gone in many directions and produced a variety of
interesting results. However, there is still no commonly agreed way for
quality management. The practice varies strongly from a concentration on
testing to a large-scale quality management process. 

 One main problem is that many of the tools and methods work rather isolated.
 Hence, quality is tackled on many levels without a combined
 strategy \cite{wagner:wosq07}. What is missing is a clear integration of these
 single efforts. One prerequisite for such an integration is a quality management
 sub-process in the overall development process. The process defines the
 roles, activities and artefacts and how these work together. A second
 prerequisite is a clear quality model that defines quality for the software to
 be developed. The term \emph{quality model} is used in various contexts but
 we use it here to denote a model that breaks down the complex concept 
 \emph{quality} and thereby makes it more concrete. These concrete descriptions
 of quality could then be used to construct, assess and predict software quality.

\subsection{Problem}
Current quality models such as the ISO 9126 \cite{iso9126-1:2001} have
widely acknowledged problems \cite{kitchenham96,deissenb:icsm07}.
Especially as a basis for assessment and prediction, the defined ``-ilities''
are too abstract. A clear transition to measurements is therefore difficult in practice.
Hence, quantitative quality assessment and prediction is usually done
without direct use of such a quality model. This, in turn, leads to isolated
solutions in quality management.

\subsection{Contribution} 

We use the previously proposed activity-based quality models (ABQM) as a basis
for quality assessment and prediction. They provide a clear structure of
quality and detailed information about quality-influences. Activity-based
quality models have proven useful in practice to structure quality and to
generate corresponding guidelines and checklists. In this paper, we
add a systematic transition from ABQMs to Bayesian networks
in order to enhance their assessment and prediction capabilities.  A 4-step approach
is defined that generates a Bayesian network using an activity-based
quality model and an assessment or prediction goal. The approach is
demonstrated in a proof of concept.

\subsection{Outline}

We first motivate and introduce activity-based quality models in Section 
\ref{sec:quality_models}. In Section \ref{sec:approach} the 4-step
approach for systematically constructing a Bayesian network from an
activity-based quality model is proposed. The approach is then
demonstrated in a proof of concept in Section \ref{sec:proof} using
publicly available data from a NASA project. Related work is discussed
in Section \ref{sec:related} and final conclusions are given in
Section \ref{sec:conclusions},

\section{Activity-Based Quality Models}
\label{sec:quality_models}

Quality models describe in a structured way the meaning of a software's 
quality. We introduce the use of general quality models and how the
modelling of activities and facts helps to define quality more precisely.

\subsection{Software Quality Models}

If quality requirements are handled at all in a software project, 
quality models will be an integral part of it.
The quality model describes what is meant by \emph{quality} and refines
this concept in a structured way. In practice, this is often reduced to
metrics such as \emph{number of defects} or high-level descriptions as
given by the ISO 9126 \cite{iso9126-1:2001}.

In general,
there are two main uses of quality models in a software project: (1) as a basis
for defining quality requirements and (2) for relating quality assurance
techniques and measurements to the quality requirements. The first is
commonly done by constraining well-known quality attributes (reliability,
maintainability,
\ldots) as defined in a quality model. In practice, this is often reduced
to simple statements such as ``The system shall be easily maintainable.''
The second use is often not explicitly considered. However, certain metrics are
nearly always measured such as the number of faults in the system 
detected by inspections and testing. The relationship between these
measures and quality attributes remains unclear. The reason lies in the
lack of practical means to define metrics for these high-level
quality attributes. Hence, more structure and more detail is needed in
quality models to integrate them closely in the development process.

\subsection{Facts and Activities}

It has been proposed to use activity-based quality models (ABQM) in order
to address the shortcomings of existing quality models \cite{deissenb:icsm07}. 
The idea is to avoid to use high-level ``-ilities'' for defining quality but to break
it down into detailed facts and their influence on activities performed on and 
with the system. In \cite{deissenb:icsm07} this was shown for 
maintainability, in \cite{winter:interactive07} for usability. In addition to information
 about the characteristics of a system,
the model contains further important facts about the process, the team and the
environment and  their respective influence on maintenance
activities such as \entity{Code Reading}, \entity{Modification}, or
\entity{Test}.
For example, redundant methods in the source code, also called
clones, exhibit a negative influence on modifications of the system because
changes to clones have to be performed in several
places in the source code. 

For ABQMs, an explicit meta-model or structure model was defined in order
to characterise the quality model elements and their relationships. Four
elements of the meta-model are most important: \entity{Entity}, 
\attribute{Attribute}, \entity{Impact} and \entity{Activity}.  An \entity{entity}
can be any thing, animate or inanimate, that can have an influence on the software's 
quality, e.g.\ the source code of a method or the involved testers. These 
entities are characterised by attributes such as \attribute{Structuredness} 
or \attribute{Conformity}. The combination of an entity and an attribute is
called a \emph{fact}. We use the notation \fact{Entity}{Attribute} for a
fact. For the example of code clones, we can write \fact{Method}{Redundancy}.
These facts are assessable
either by automatic measurement or by manual review. If possible, we define
applicable metrics for measuring the facts inside the ABQM.

An influence of a fact is then specified by an \entity{Impact}. We concentrate
on the influences on activities, i.e.\ anything that
is done with the system. For example, \entity{Maintenance} or \entity{Use}
are high-level activities. The impact
on an \entity{Activity} can be positive or negative. 
We complete the code clone example by adding the impact on 
\entity{Modification}: \impactn{Method}{Redundancy}{Modification}.
This means that if a system
entity \entity{Method} exhibits the attribute \attribute{Redundancy} it will
have a negative impact on the \entity{Modification} activity, i.e.\ changing
the method. 
A further example is the following tuple that
describes consistent identifiers:
\impactp{Identifier}{Consistency}{Modification}
It means that identifiers that can be shown to be consistent have a positive
influence on the modification activity of the maintainer of the system. In the
model itself, we document more information such as textual descriptions,
sources, assessment descriptions and so on. However, the short notation captures
the essential relationships.

\begin{figure}[htb]
\centering
\includegraphics[width=0.45\textwidth]{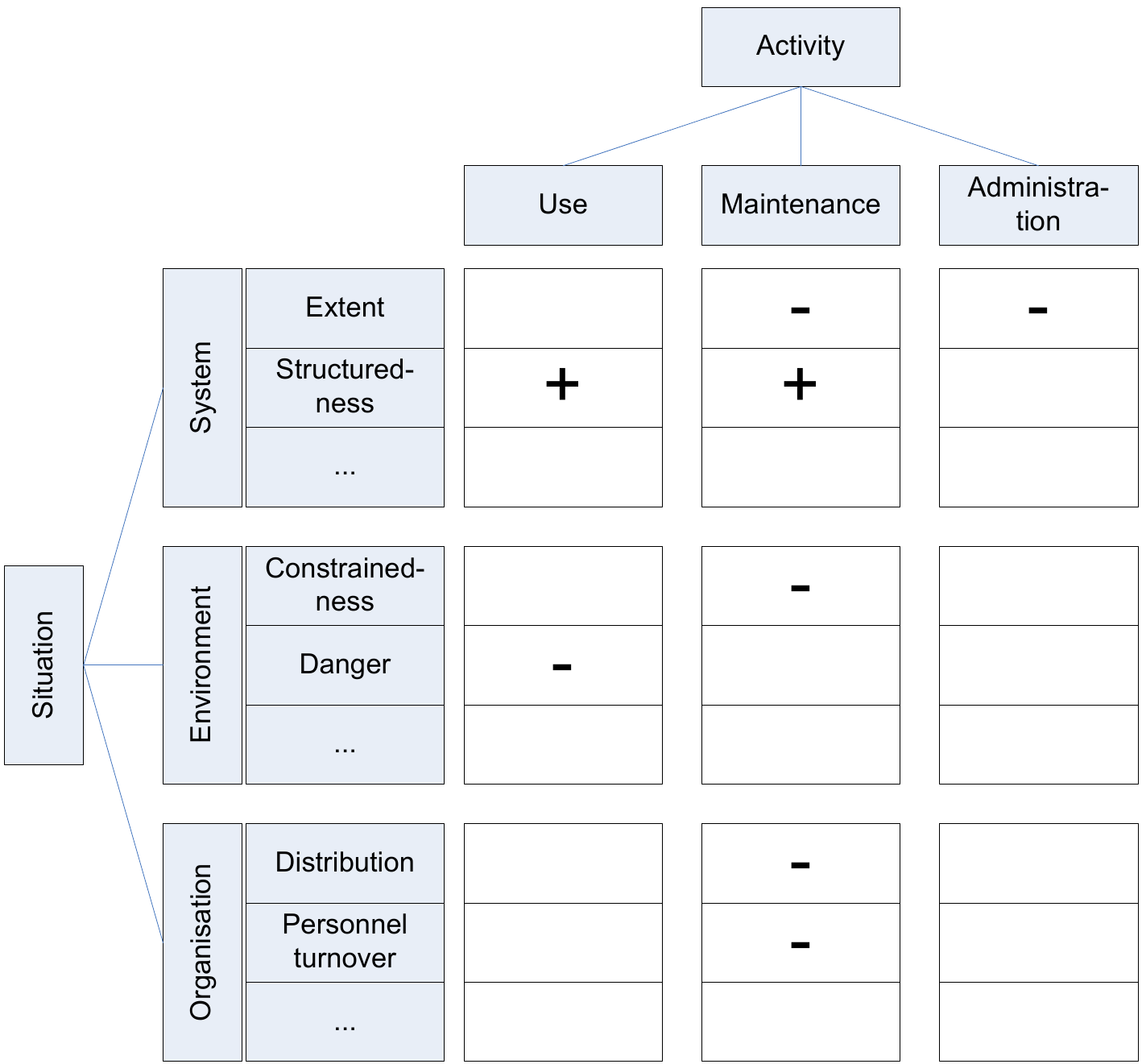}
\caption{High-level view on an activity-based quality model as a matrix}
\label{fig:quality_matrix}
\end{figure}

The model does not only contain the impacts of facts on activities but
also the relationships among these. Facts as well as activities are
organised in hierarchies. A top-level activity \entity{Activity} has
sub-activities such as \entity{Use}, \entity{Maintenance} or
\entity{Administration}. These examples are depicted in Figure
\ref{fig:quality_matrix}. In realistic quality models, they are then further
refined. For example, maintenance can have sub-activities such as 
\entity{Code Reading} and \entity{Modification}.

Because facts are a composite of an \emph{entity} and an \emph{attribute},
the organisation in a hierarchy is straightforward. Hierarchical relationships
between entities do usually already exist. The top-level in Figure 
\ref{fig:quality_matrix} is the \entity{situation} of the
software development project. It contains, for example, the \entity{System},
its \entity{Environment} and the development \entity{Organisation}. Again, these
entities need to be further refined. For example, the system consists of
the source code as well as the executable.
All entities can be described with attributes such as the
\attribute{Structuredness} of the \entity{System}. 

The two hierarchies, the fact tree and the
activity tree, together with the impacts of the facts on the activities can
then be visualised using a matrix as in Figure \ref{fig:quality_matrix}. The 
fact tree is shown
on the left, the activity tree on the top. The impacts are depicted by entries
in the matrix where a ``+'' denotes a positive and a ``-''
a negative impact.

The associations between facts in the fact tree can have two different meanings.
Either an entity is a part or a kind of its super-entity. Along the 
inheritance associations, parts and attributes are inherited. Hence, it allows
a more compact description and prevents omissions in the model. For example,
naming conventions usually are valid for all identifiers no matter whether
they are class names, file names, or variable names.

Having defined all these entries in the ABQM, we can define which
activities we want to support and which influencing facts need to be
defined. In terms of the above example, if we want to support the activity
\entity{Modification}, we know that we need to inspect the identifiers for
their consistency. 

There exists a prototype tool to define this kind of large and detailed
quality models \cite{deissenb:icsm07}. Besides the easier creation and
modification of the model, this has also the advantage that we can automate
certain quality assurance tasks. For example, by using the tool we can
automatically generate customised review guidelines for specific views.

\section{Assessment and Prediction\\ Approach}
\label{sec:approach}

Although activity-based quality models have proven to be useful in
practice, there is no systematic approach for using such quality models
in measuring. Hence, there are no quantitative assessments and
predictions possible so far. We propose an approach that can be used
for systematically deriving assessment and prediction models from a
activity-based quality model.

\subsection{Aim and Basic Idea}

The general aim of the approach is to provide quality managers with
a systematic method to derive assessments and predictions from an
activity-based quality model. In the ABQM, there are definitions
of what quality means w.r.t.~different situations, artefacts, and considered
activities. Currently, we give a textual description in the quality model
how a fact could be assessed. For example, the fact described by
\fact{Method}{Redundancy}
contains the following assessment description: ``This fact can be assessed
manually or semi-automatically. For the automatic assessments there are
tools such as CloneDetective or CCFinder to detect redundant parts of the
source code.'' This information is interesting for quality assurance planning
but cannot be directly used for an overall assessment, let alone prediction.

Moreover, as the basic principle of activity-based quality models is that the
most important question in quality is how well activities can be performed on
and with the system, not only facts but also activities should be assessed and predicted
from the knowledge of facts and impacts. Currently, only the qualitative
statement whether an impact is positive or negative is made. This is suitable
for rough assessments only. For more comprehensive and precise assessments
of the current state and prediction of future states, a more sophisticated approach
is needed. It has to systematically help to use the given relationships and enrich
them with quantitative information.

As most facts and especially the relationships between facts and activities
have an associated uncertainty, statistical methods are needed. Usually we
cannot determine the exact relationship but can derive an uncertain range. Also
values measured can be uncertain, e.g.\ values from expert opinion. Moreover,
a statistical method is needed that can directly model the dependencies of
different factors from the quality model. We identified Bayesian networks as
most suitable for that task.

\subsection{Bayesian Networks}

Bayesian networks, also known as Bayesian belief nets or belief
networks, are a modelling technique for causal relationship based on
Bayesian inference. They are represented as a directed acyclic graph (DAG)
with nodes for uncertain variables and edges for directed relationships
between the variables. This graph models all the relationships abstractly.

For each node or variable there is a corresponding \emph{node probability 
table} (NPT). These tables define the relationships and the uncertainty of
these variables. The
variables are usually discrete with a fixed number of states. For each
state, the probability that the variable is in this state is given. If
there are parent nodes, i.e.\ a node that influences the current node,
these probabilities are defined in dependence on the states of these 
parents. An example is shown in Table~\ref{tab:example_npt}. There the
variable is with a probability of 60\% in the state \emph{true}
if both parents are in the state \emph{low}, and with 45\% in \emph{true}
if the first parent is in \emph{high} and the second is in \emph{low}.

\begin{table}[htb]
  \caption{An example NPT for a variable with two states and two parents
          \label{tab:example_npt}}
\begin{center}
\begin{tabular}{lrrrrrr}
\hline
 & & low & & & high & \\
\hline
 & low & med & high & low & med & high\\
\hline
true & 0.6 & 0.65 & 0.3 & 0.45 & 0.23 & 0.05\\
false & 0.4 & 0.35 & 0.7 & 0.55 & 0.77 & 0.95\\
\hline
\end{tabular}
\end{center}
\end{table}

The process of building a Bayesian network contains the identification
of interesting variables that shall be modelled, representing them as nodes,
constructing the topology and constructing the NPTs. Each of these steps
is important and non-trivial. First, the identification of \emph{interesting}
variables includes the assumption that the model builder can decide on
some basis what is interesting. In many cases, this is not clear beforehand.
One possibility is to include many variables and use sensitivity analysis to
remove insignificant variables.

Second, the creation of the topology uses the assumption that the model
builder can decide on the dependence and independence of the
identified variables. In the process of building the Bayesian network,
especially for independence assumptions (i.e.\ missing edges in the
graph) detailed justifications should be given. Third, the problem of
constructing NPTs is widely acknowledged in the literature \cite{fenton:tr07}. It involves
defining quantitative relationships between variables. There are various
possible methods for this quantification such as a probability wheel or
regression from empirically collected data. All methods have their pros and
cons.

It is important to note that each of these steps is important and errors
in each of these steps can have a large effect on the outcome. Bayesian
networks and the corresponding tool support make it easy to build models
and get quantitative results. However, one needs to be aware that many
assumptions are embedded in a Bayesian network that need to be validated
before it can be trusted.

\subsection{Four Steps for Network Building}

We propose a four-step approach for building a Bayesian network
as assessment and prediction model derived systematically from an
activity-based quality model. The resulting Bayesian network contains three
types of nodes:
\begin{itemize}
\item \emph{Activity nodes} that represent activities from the quality model
\item \emph{Fact nodes} that represent facts from the quality model
\item \emph{Indicator nodes} that represent metrics for activities or facts
\end{itemize}
We need four steps to derive these nodes from the information of the
ABQM. First, we identify the relevant activities with indicators
based on the assessment or prediction goal. Second, influences by
sub-activities and facts are identified. This step is repeated recursively
for sub-activities. The resulting facts together with their impacts are modelled.
Third, suitable indicators for the facts are added. Fourth, the node probability
tables (NPT) are defined to reflect the quantitative relationships. Having that,
the Bayesian network can be used for simulation by setting values for any
of the nodes. However, we first describe the four steps in more detail.

The first step is a goal-based derivation of relevant activities and their indicators.
We use GQM \cite{basili94} to structure that derivation. We first define the
assessment or prediction \emph{goal}, for example, optimal maintenance
planning or optimisation of the security assurance. The goal shows the relevant
activities, such as \emph{maintenance} or \emph{attack}. This is refined by
stating \emph{questions} that need to be answered to reach that goal. For example, 
will the maintenance effort increase over the next year or how often will there
be a successful attack in the next year? Finally, we derive \emph{metrics} or
indicators that allow a measurement to answer the question. In the examples,
that can be effort change trend or number of harmful attacks over the next year.

\begin{figure*}[htb]
\centering
\includegraphics[width=\textwidth]{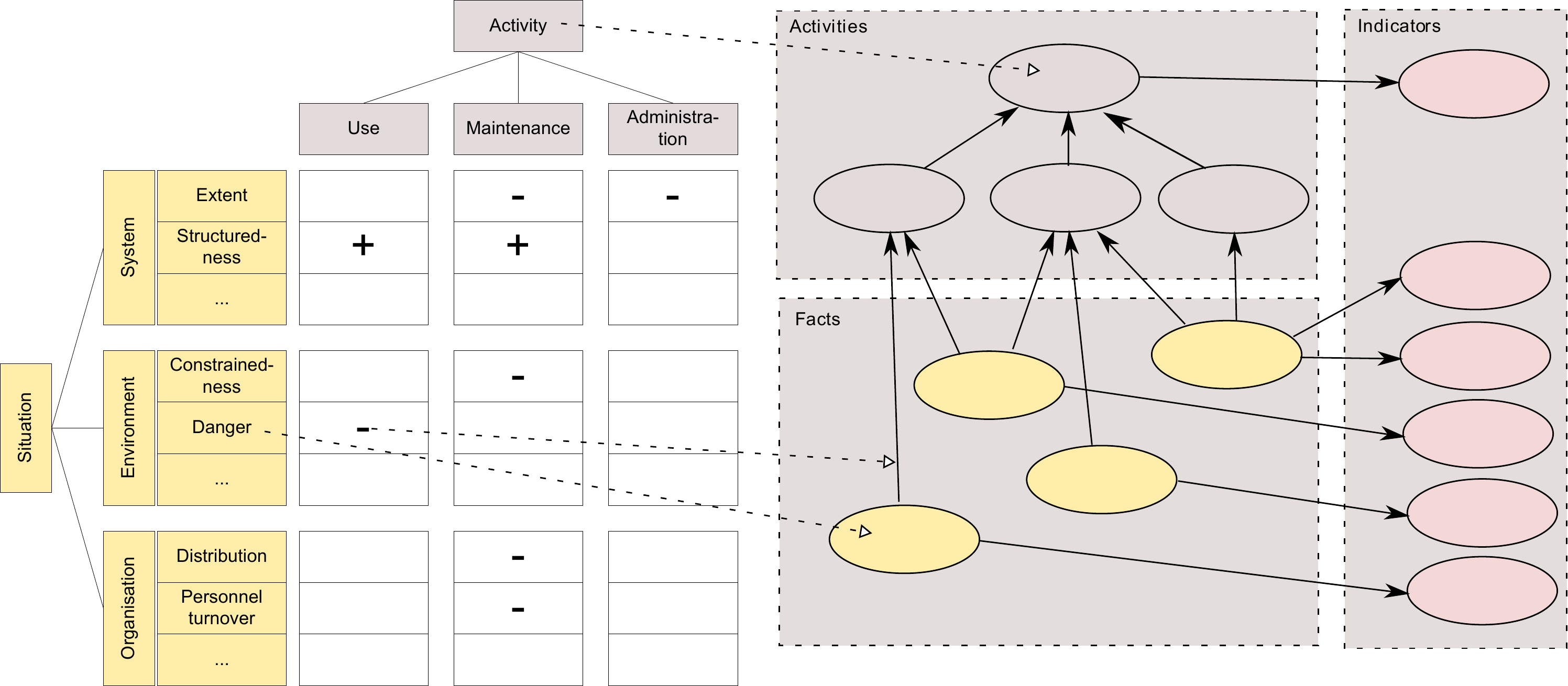}
\caption{Mapping of quality model elements to Bayesian network elements.
               The dashed arrows indicate examples for mappings.}
\label{fig:qm2bbn}
\end{figure*}

In the second step, we use the quality model to identify the other factors that
are related to the identified activities. There are two possibilities: (1) there are
sub-activities of the identified activities and (2) there are impacts from facts to
the identified activities. We repeat this recursively for the sub-activities until
all facts that have an impact on the activities sub-tree below the
identified activity are collected. For each activity we immediately see the impacts and hence
the corresponding facts. All activities and facts identified this way are modelled
as nodes in the Bayesian network. We add edges from sub-activities to
super-activities and from facts to activities on which they have an impact.
Figure \ref{fig:qm2bbn} gives an abstract overview of that mapping from the
quality model to the Bayesian network.

In the third step, we add additional nodes as indicators for each fact and
activity node that we want a measurement for. The indicator for our relevant
activity was defined in the first step. Hence, we can add additional indicators
for sub-activities if needed. In any case, there need to be at least one indicator
for each fact that is modelled. There might be a precise description in the
quality model already. Otherwise, we need to derive our own metric or use an
existing one from the literature.
The edges are directed from the activity and fact nodes to the indicators, i.e.\ the
indicators are dependent on the facts and activities. An indicator is only an expression
of the underlying factor it describes.

A main advantage of using a ABQMs as a basis for the Bayesian network
is that it prescribes its topology. One of the main points of such quality models
is to qualitatively describe the relationships between different factors that
are relevant for software quality. We rely on that and assume that all
dependencies have been modelled and that all other factors are independent.
On the one hand, this constrains the validity of the results of the Bayesian network by the
validity of the ABQM. On the other hand, it frees the network builder from
reasoning about independence and dependence.

Finally, the fourth step, enriches the Bayesian network with quantitative information.
This includes defining node states as well as filling the NPT for each node. The
activity and fact nodes are usually modelled as \emph{ranked} nodes, i.e.\ in a
ordinal scale. The most common example is low, medium, and high.
This has advantages in evaluation and aggregation. The evaluation
is easier as not precise numbers have to be determined but coarse-grained levels.
These levels actually reflect much more the high uncertainty in the data. Also in
aggregating over nodes (up the hierarchy in the activity tree) coarse-grained ranked
data is more simple to handle. For only a few levels in ordinal scale, it is easier
to define a aggregation specification than for continuous data. To define the NPT,
we use an approach proposed by Fenton, Neil and Galan Caballero \cite{fenton07a}.
The basic idea is to formalise the behaviour observed with experts that have to
estimate NPTs. They usually estimate the central tendency or some extreme values
based on the influencing nodes. The remaining cells of the table are then filled
accordingly. This is similar to linear regression where a Normal distribution is
used to model the uncertainty. We use the doubly truncated Normal distribution
that is only defined in the $[0,1]$ region. It allows to model a variety of shapes
depending on the mean and variation. It allows, for example, to model the NPT
of a node by a weighted mean over the influencing nodes.

The node states of indicator nodes depend on the scale of the indicator used.
This often will be continuous or discrete interval states such as lines of codes
in intervals of a hundred. The NPTs of the indicator nodes
are then defined using either common industry distributions or information from
company-internal measurements. For example, typical LOC distributions can be
accumulated over time. The influence of the activity or fact node, respectively,
it belongs to can be modelled in at least two ways: (1) partitioned expressions
and (2) arithmetic expressions. The latter describes a direct arithmetical
relationship from the level in the activity or fact node to the indicator. Using
a partitioned expression, the additional uncertainty can be expressed by
defining probability distributions for each level of the activity or fact node.

\subsection{Usage of the Bayesian Network}

A main feature of Bayesian networks is their capability to simulate different
scenarios. Having built the Bayesian network based on the activity-based quality
model, we can ask ``what if?'' questions. These questions are formulated in
scenarios that can be simulated and compared. A scenario involves adding
additional information to the model, more specifically, adding an observation for
a node. This way, the uncertainty is removed and the consequences for the other
nodes can be calculated. Interestingly, in a Bayesian network it is possible to do
forward as well as backward inference, i.e.\ information can be added to any node
and the effect is calculated in any direction of the graph.

A first straightforward scenario is to add the currently measured values for the
fact indicator nodes. This will drive the calculation up to the activities and the
activity indicator node. This indicator node then shows the probability distribution
for its value, i.e. the value of the activity. Then changes can be made to the
fact indicator nodes in other scenarios to reflect possible changes. Their effect
on the activity can be predicted. Another interesting scenario is to set a desired
value for the activity indicator and let the network calculate the most probable
explanation in the fact indicators. It shows the values that should be reached
in order to fulfil the goal.

\section{Proof of Concept}
\label{sec:proof}

The above presented approach can be used in many different contexts and
to answer various assessment and prediction questions. We provide a
proof of concept here that applies the approach to a publicly available data
set.

\subsection{Goal}

Currently, we cannot provide a full validation of the assessment and prediction
approach because this would involve measuring a large number of facts from
the quality model in order to take full advantage from the knowledge contained
in it. This data is not available in public data sets. Measuring this at a company
will need time and effort. Only then a sensible analysis of the predictive validity
and comparisons with other prediction models are possible.

Nevertheless, we provide a proof of concept application of the approach
based on a small extract of our quality model for maintainability for which there
is publicly available data. It demonstrates the basic principles of the approach
on a real data set. This way, we can analyse whether the approach is
feasible in an almost realistic setting. The predictive validity in the proof of
concept can give an indication of the usefulness of the approach.
There should at least be an improvement over industry average values.

\subsection{Context}

The proof of concept comprises the prediction of the maintainability of a
system. The system under analysis was developed by NASA in the project
CM1 for which the data has been publicised \cite{nasa-mdp}. The system
itself is an instrument in a space craft developed in C. Various metrics,
especially the McCabe and Halstead metrics, have been collected in this
project.

As the quality model, we use the activity-based quality model for maintainability
from \cite{deissenb:icsm07}. It contains a complete activity tree for maintenance
as well as about 200 facts with an impact on those activities. The CM1
data set does not contain data for all of these facts but we choose a small set
of facts and corresponding activities to predict maintainability. The choice was
therefore guided by the availability of the data. The actual maintainability can be
judged for CM1 because the effort for several changes has been documented.
We will use this as our surrogate measure.

For modelling the Bayesian network, we use the tool 
AgenaRisk\footnote{\url{http://www.agenarisk.com/}}. It provides
a complete tool environment for Bayesian networks including the usage of
expressions for describing the NPT and sensitivity analysis.

\subsection{Procedure}

The first step in our assessment and prediction approach is to identify the relevant
activities and corresponding indicators (metrics) in a GQM-like procedure. We
want to analyse maintainability and we assume that the quality manager is
interested in the question of how high the maintenance efforts will be in the
future. This information helps in planning the maintenance team. Hence, we
formulate the \textbf{Goal} as ``Planning of future maintenance efforts''. We
can directly identify the activity \entity{Maintenance} in that goal. To further
operationalise that, we define the corresponding \textbf{question} as ``What
will be the maintenance effort per change request?''. This information, probably
together with a prediction of the yearly number of change requests, would give
a good basis for maintenance planning. However, we concentrate only on the
question about the effort per change request. This leads us straightforwardly
to the \textbf{metric} ``average effort per change request''. Hence, an analysis
of the mean is needed.

\begin{figure*}[htb]
\centering
\includegraphics[width=\textwidth]{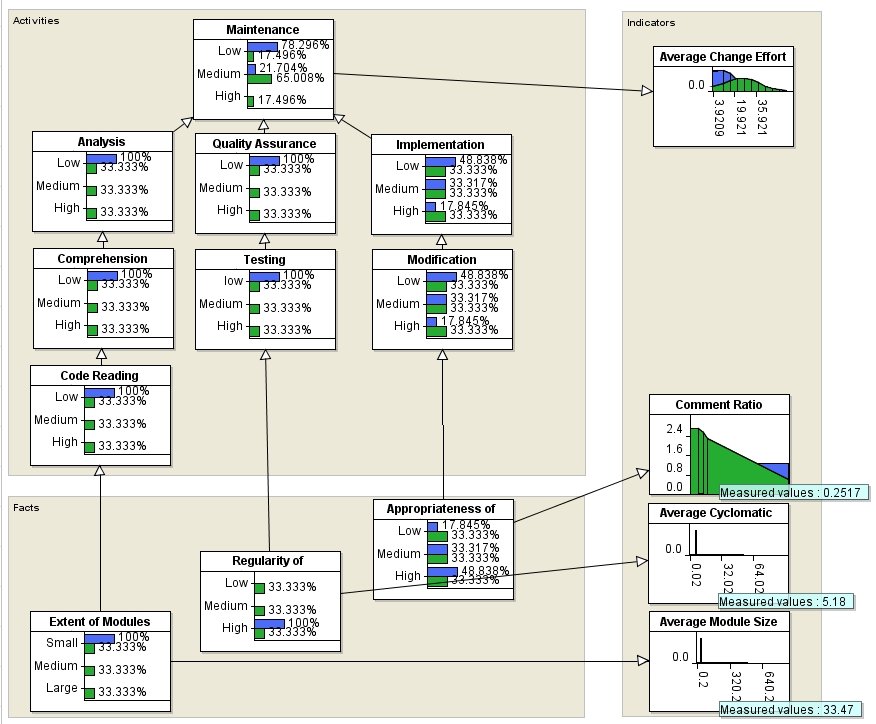}
\caption{The resulting Bayesian network}
\label{fig:screenshot}
\end{figure*}

In the second step, we already start building the Bayesian network. We look
at the maintainability model and find no impacts directly on \entity{Maintenance}
that should be considered here. However, we find 10 sub-activities including the
following 3 that we will further analyse: \entity{Quality Assurance}, \entity{Implementation}
and \entity{Analysis}.  All three have impacts from facts but having in
mind the available data, we ignore these and use the further sub-activities
\entity{Testing}, \entity{Modification} and \entity{Comprehension} with its child
\entity{Code Reading}. We create nodes for these activities and connect them with
edges corresponding to their hierarchy. They can be seen in Figure \ref{fig:screenshot}
in the box ``Activities''.

We only include 3 impacts on each of the lowest level activities chosen so far
because we can find corresponding data in the CM1 data set. These facts together
with their impacts are:
\begin{itemize}
\item
\impactn{Module}{Extent}{Code Reading}\\
The size of a module has an impact on reading the code of the module. In essence,
the larger the module, the longer it takes to read it.

\item
\impactp{Implementation}{Regularity}{Testing}\\
An implementation is regular if it does not use unnecessarily nested branches. This
complex structure would render coverage by tests more difficult.
\item
\impactp{Comment}{Appropriateness}{Modification}\\
Comments need to appropriately describe the code it is associated with.
\end{itemize}

The facts are added as nodes in the Bayesian network. In Figure \ref{fig:screenshot}
they are contained in the box ``Facts''. The impacts are included as the arrows
from the facts to the activities.

The indicators that are identified in the fourth step of the approach are taken from
the available data of the data set. We only use 1 indicator per fact although we are
aware that each fact has more aspects that should be covered. For the \emph{Extent
of Modules}, we use the indicator \emph{Average module size} given in LOC.
The \emph{Regularity of the Implementation} is indicated by the \emph{Average
Cyclomatic Complexity}. This is not a particularly good indicator as it only gives a
number for the decision points in the implementation. A manual review can far better
decide whether the implementation is regular. However, we have no access to review
results. A similar reasoning holds for the indicator \emph{Comment Ratio} for the fact
\emph{Appropriateness of Comments}. The proportion of the comments in relation to
the other code is only of minor importance in comparison with the semantic appropriateness.
However, we do not have access to such a semantic judgement. The indicators can
be found in Figure \ref{fig:screenshot} on the right-hand side in the box ``Indicators''.

A difficult problem in general with Bayesian networks is the definition of the node
probability tables (NPT)  \cite{fenton:tr07}. Various method can be used to define
these NPTs that we will not describe here in detail. For the approach, we can simplify
the problem to 2 cases: (1) activities and facts NPTs and (2) indicator NPTs. The NPTs
for the activities and facts can, unless there is additional knowledge, be assumed as
uniformly distributed. The impacts are only modelled as direct influence. Here, a
special method for ranked nodes \cite{fenton07a} can be applied that simplifies the
task. For the indicator NPTs either empirically investigated distributions of the 
company or industry average distribution should be used. This forms the distribution
of the indicator values under uncertainty without any observations.

For the average effort of a change, we refer to \cite{wagner:isese06} that gave an
a mean defect removal cost of 27.4 person-hours with minimum 3.9 and maximum
66.6. Although a change does not always have to be a defect removal, it is precise
enough for the proof of concept. For the distributions of the other indicators, no published
distributions were available. Hence, our own expert opinion was used as a basis.


\subsection{Results and Discussion}

Figure \ref{fig:screenshot} shows two scenarios. The green values describe the
scenario in which no observations have been made, i.e.\ the general scenario. The
blue values are for the scenario \emph{Measured values} in which for the three fact
indicators the real measured values are set as observations. More specifically, the
comment ratio is set to .2517, the average cyclomatic complexity to 5.18 and the
average module size to 33.47 LOC. This has effects on the rating of the facts in
turn on the activities. The most interesting value, however, is the average change
effort. In the general scenario, this variable has a mean value of 27 with a standard
deviation of 12.1. The additional information about the measured values shifts the
distribution to the left. The mean decreases to 15.9 with standard deviation 8.5. This
is still not the really measured value of 6 but much closer than the industry standard
of 27. Hence, it is an improvement in comparison to just using standard numbers.

The reason for the large difference between the prediction and the real value
is probably three-fold. First, the distribution from \cite{wagner:isese06} might 
not be appropriate for the NASA environment in which the effort for a change
request seems significantly smaller than average. Second, it might also be the
case that the data in the CM1 data set does not use exactly the same measures as
in \cite{wagner:isese06}. The degree to which additional efforts for re-inspection
and re-testing are included could vary. Third, several more factors than the 3
considered might have an influence on the maintenance effort. The quality
model contains many explanations in terms of facts that should be investigated. 

Nevertheless, the point of the proof of concept was not mainly to show predictive
validity but to investigate the general feasibility. We saw that it is possible with
reasonable effort to build a Bayesian model. The 4-step approach gives direct
guidance for most of the network building. Only setting up the NPTs is still a
challenge. There are usually several possibilities how a relationship can be
expressed and with how much uncertainty it should be afflicted. This still needs
expert opinion and experimentation. Nevertheless, AgenaRisk provides very
good tool support to find easier ways to define an NPT. As our ABQMs can get
very large with a few hundreds of model elements, it remains to be evaluated
whether the approach scales when the quality model is fully mapped to a
Bayesian network. Probably, a selection of a subset of the quality model is
necessary first.


Furthermore, it is important to note that a more in-depth validation of a
resulting Bayesian network is necessary in order to ensure that all parts,
topology, node states, and NPTs, represent the interdependencies of
the quality factors good enough so that a valid statement about the quality
of the software system can be made. This is not covered by this proof-of-concept
but has to be the next step.

\section{Related Work}
\label{sec:related}

The basic idea to use Bayesian networks for assessing and predicting software
quality has been developed mainly by Fenton, Neil and Littlewood. They
introduced Bayesian networks as a useful tool and applied it in various
contexts related to software quality. In \cite{fenton99} they formulate a critique
on current defect prediction models and suggest to use Bayesian networks.
Other researchers also used Bayesian networks for software quality prediction
similarly \cite{amasaki05,perezminana06}

The work closest to the approach proposed in this paper is \cite{neil99}.
They discuss quality models such as the ISO 9126 \cite{iso9126-1:2001}
and their problems such as the undefinedness of the relationships in such
a model. They aim at solving these problems by defining Bayesian networks
for quality attributes directly. Our work differentiates in using a defined structure
for quality models that contain far more details as common quality models.
This structure and detail allows a straightforward derivation of a Bayesian
network from the quality model. This has the advantage that the basic quality
model can also be used for other purposes then prediction such as the
specification of quality requirements.

Beaver, Schiavone and Berrios \cite{beaver05} also used a Bayesian network
to predict software quality including diverse factors such as team skill and
process maturity. In his thesis \cite{beaver06}, Beaver even compared the
approach to neural networks and Least Squares regressions that both were
outperformed by the Bayesian network. However, they did not rely on a
structured quality model as in our approach.

\section{Conclusions}\label{sec:conclusions}

A high goal in software quality management is the reliable quantitative
assessment and prediction of software quality. Many efforts
in building assessment and prediction models have given various insights
in the usefulness but also the constraints of such models. However, these
models have not been tightly integrated into other quality management
activities. Activity-based quality models have proven in practice to be a
solid foundation for defining quality on a detailed level. However, quantitative
analyses have not been directly possible so far.

Bayesian networks have been shown to provide promising results in
quality predictions. Because of that and their clear structuring that can
straightforwardly reflect the structure of activity-based quality models, a
4-step approach for transferring activity-based quality models to Bayesian
networks was proposed. It allows to systematically construct a Bayesian
network that uses the knowledge encoded in the quality model to provide
information about a given assessment or prediction goal. In the terminology of
\cite{deissenb:wosq09}, we use a \emph{quality definition model} (the ABQM)
and enrich it with a \emph{quality assessment} and \emph{quality prediction
model} (the BN).

Although not fully validated, we demonstrated the approach in a case
study using real NASA project data. The proof of concept showed the
applicability of the approach on such a project and even could improve
the prediction in comparison to industry standard values. The prediction
was not very accurate but that can have various reasons such as a
different effort distribution at NASA or the influence of more factors that
have not been considered because there was no data available.

The use of Bayesian networks opens many possibilities. Most interestingly,
after building a large Bayesian network, a sensitivity analysis of that
network can be performed. This can answer the practically very relevant
question which of the factors are the most important ones. It would
allow to reduce the measurement efforts significantly by concentrating
on these most influential facts.

We plan to apply this approach in future case studies at our industrial
partners in order to further validate the approach. A comprehensive
analysis of the predictive validity is necessary to judge the usefulness
of the approach and to compare it with other means for assessment
and prediction.

\section{Acknowledgments}

This work has partially been supported by the German Federal Ministry of Education and Research (BMBF) in the project QuaMoCo (01 IS 08023B).

\balance
\bibliographystyle{abbrv}

\end{document}